\documentclass[aps,superscriptaddress,amsmath,amssymb,floatfix,showpacs,notitlepage,twocolumn]{revtex4-1}
\usepackage{graphicx,graphics,setspace,epsfig,color}
\usepackage[letterpaper, dvips,width=7.5in,height=8.5in,includemp=false]{geometry}
\usepackage[vcentermath]{}
\usepackage{epsf}
\usepackage{amscd}
\usepackage{amsmath}

\newcommand{\beq}{\begin{equation}}
\newcommand{\eeq}{\end{equation}}
\newcommand{\bea}{\begin{eqnarray}}
\newcommand{\eea}{\end{eqnarray}}
\newcommand{\nn}{\nonumber}
\newcommand{\eq}[1]{eq.~(\ref{#1})}

\newcommand{\threeptwo}{${}^3P_2\ $}

\begin{document}

\title[title]{Goldstone modes in the neutron star core}
\author{ Paulo F. Bedaque }
\affiliation{Department of Physics, University of Maryland, College Park, MD 20742}
\author{ Sanjay Reddy}
\affiliation{Institute for Nuclear Theory, University of Washington, Seattle, WA}


\begin{abstract}
We formulate a theory of Goldstone bosons and their interactions in the superfluid and  superconducting phase of dense nucleonic matter at densities of relevance to the neutron star core. For typical neutron star temperatures in the range $T=10^{6}-10^{9}$ K,  the Goldstone mode associated with  rotational symmetry, called {\em angulons}, couple weakly to each other and to electrons. Consequently, these modes have anomalously large mean free paths and can contribute to both diffusive and ballistic transport of heat and momentum. In contrast, the two Goldstone bosons associated with density oscillations of the neutron and electron + proton fluids, called {\em superfluid phonons}, mix and couple strongly to electrons. They have shorter mean free paths, and their contribution to transport is negligible. Long-wavelength {\rm superfluid phonons} and {\em angulons} can play a role in neutron star seismology, and lead to interesting phenomenology as angulons couple to magnetic fields and have anisotropic dispersion relations.    
\end{abstract}

\maketitle
It is likely that neutron-rich matter encountered in the neutron star core is both superfluid and superconducting. At the densities realized in the core, attractive s-wave interactions between protons, and p-wave interactions between neutrons naturally leads to the formation of Cooper pairs of nucleons at the Fermi surface. Neutrons are expected to form a $^3P_2$ superfluid and protons form a $^1S_0$ superconductor. Model calculations indicate that the typical pairing energy or equivalently the energy gap, denoted by $\Delta  $ are small ($\Delta \lesssim 1$ MeV $ \simeq k_B 10^{10}$ K) compared to the Fermi energy, but are large compared the typical temperatures $T\simeq 10^7-10^8$ K encountered in old and cold neutron stars. Hence, it has been known for some time that pairing will not affect the equation of state, but can dramatically alter key transport and cooling properties of neutron stars.  

When $k_B~T \ll \Delta$, the excitation of neutrons and protons is  exponentially suppressed by the factor $\exp{(-\Delta/k_BT)}$ and the only relevant excitations are electrons and Goldstone bosons (GBs) associated with the symmetry breaking in the superfluid and superconducting ground states. 
Since nucleon excitations are gapped, the theoretical description of transport properties are dominated by GBs and electrons, and we show here that they are weakly interacting and that kinetic theory applies. As a consequence, long-wavelength oscillations, shear viscosity and thermal conductivity, which play a role in interpreting various transport phenomena in neutron stars, can be calculated systematically in terms of just a handful of low-energy constants. 

The letter is organized as follows. First, we use general symmetry arguments and simple dynamical considerations to discuss the structure of the low energy theory of GBs. Then, we use this effective theory to calculate the dispersion relations and mean free paths of all the GBs to show that some of them are especially weakly coupled. We conclude by estimating neutron star oscillation frequencies, and the GB mode contribution to thermal conductivity and shear viscosity, and  discuss its implications for neutron stars.  Through out this letter, unless we explicitly note otherwise, we use natural units with $\hbar=1,c=1$.               

At the high densities encountered in the core, s-wave interactions between neutrons is repulsive and attractive p-wave interactions favor spin-one Cooper pairs with angular momentum in the $^3P_2$ channel. The resulting condensate, which we discuss in more detail below, breaks rotational invariance as well as the global $U(1)$ symmetry associated with rotations of the phase of the neutron wave function. Because the proton fraction is smaller, typically less than $10\%$, s-wave interactions between protons remain attractive even in the core, and this leads to the formation of spin-zero proton Cooper pairs in the $^1S_0$ channel. The resulting superconducting condensate breaks the $U(1)$ symmetry associated with rotations of the phase of the proton wave function.  In contrast, due to their large Fermi energy ($\gtrsim100$ MeV) electrons form an nearly ideal, relativistic and degenerate Fermi gas. 

In the neutron superfluid there are Goldstone modes associated with breaking of rotational symmetry by the $^3P_2$ condensate aptly labelled as  {\em angulons} in \cite{Bedaque:2003wj}, and one {\em superfluid phonon} mode associated with the neutron number $U(1)$ symmetry. Naively, one may expect that the Goldstone mode associated with the proton superconductor will not be massless due to long-range Coulomb interactions. However, due to efficient electron screening, a massless Goldstone does exist, and it corresponds to a charge neutral oscillation of the proton condensate and the electron fluid \cite{Baldo:2011iz}. We now discuss in detail the theory needed to describe the propagation of these massless modes.   

The neutron condensate is a spin-2 object which is given by 
\beq
\langle N^T  \sigma_2 \sigma^i \overleftrightarrow{\nabla}^j  N\rangle = \Delta^0_{ij}~e^{i\phi},
\eeq 
where $\Delta^0_{ij}$ is a symmetric traceless tensor and $\phi$ is a scalar.  Different symmetric traceless tensors break the rotation group in different ways so there are several possible \threeptwo phases.     
Around the critical temperature one can rely on BCS and strong coupling  estimates of the parameters of the Ginsburg-Landau free energy to conclude that the ground state is of the form $ \Delta^0_{ij}\sim {\rm  diag}(1, 1, -2)$ (or, of course, any rotation of this matrix)\cite{Sauls:1978,Richardson:1972xn}. The structure of the gap equations are such that, at least within the BCS framework, the relative order of the different states is not changes as temperature, density or microscopic interactions change \cite{Khodel:1998hn} so it is reasonable to assume that the ground state of neutron matter is in a phase characterized by the  $ \Delta^0_{ij}\sim {\rm  diag}(1, 1, -2)$ form of the condensate. This will be an assumption underlying our analysis although many of our qualitative conclusions are independent of it.

The presence of the condensate $ \Delta^0_{ij}\sim {\rm  diag}(1, 1, -2)$ breaks spontaneously the symmetry of the system under rotations, except for those around the $z$-axis. Thus, as first realized in \cite{Bedaque:2003wj} we expect the presence of two gapless excitations above the ground state, named ``angulons", corresponding to rotations of the condensate around the $x$ and $y$ axis. Angulons were then studied in more detail in \cite{Bedaque:2012} where, with mild assumptions, their properties were quantitatively estimated.

 
These properties are succinctly encapsulated is the lagrangian given by  
\bea\label{eq:L_ang}
\mathcal{L} _{\rm ang}
&=&  \sum_{i=1,2} \left[
\frac{1}{2} (\partial_0\beta_i)^2 
-\frac{1}{2} 
{v_\perp^i}^2 ((\partial_x\beta_i)^2 + (\partial_y\beta_i)^2)   
+
v_\parallel^2  (\partial_z\beta_i)^2 
\right]\nn\\
&+&
\frac{e g_n f_\beta}{2M\sqrt{-\nabla^2_\perp}}
\left[
\mathbf{B}_1\partial_0 ( \partial_y \beta_1+ \partial_x \beta_2)
+
\mathbf{B}_2\partial_0 ( \partial_x \beta_1- \partial_y \beta_2)
\right]\nn\\
& +& \mathcal{O}\left(\beta^2 \frac{  (\partial\beta)^2}{f_\beta^2}\right),
\eea where
\bea\label{eq:v_and_f}
{v_\perp^1}^2 = \frac{24\pi\sqrt{3}}{18(9+\pi\sqrt{3})} v_F^2, &\qquad &{v_\perp^2}^2 = \frac{81-4\pi\sqrt{3}}{18(9+\pi\sqrt{3})} v_F^2\nn\\
{v_\parallel}^2 = \frac{81-2\pi\sqrt{3}}{18(9+\pi\sqrt{3})} v_F^2, &\qquad&   f_\beta^2 = \frac{Mk_F}{\pi^2}~\left(1+\frac{\pi}{3\sqrt{3}}\right),
\eea  $g_n\approx-1.91$ is the neutron magnetic moment in units of the nuclear Bohr magneton, $\mathbf{B}$ is the magnetic field, $k_{Fn}$ the neutron Fermi momentum, $M$ the nucleon mass, $v_F=k_{Fn}/M$ is the neutron fermi velocity, and $e=\sqrt{\alpha_{em}/4\pi^2}$  the electron charge. The values in \eq{eq:v_and_f} receive Fermi liquid corrections not yet computed. The fields $\beta_{1,2}$ are linear combinations of the fields describing rotations of the condensate around the $x$ and $y$ axis which mix among themselves; in terms of the original fields the lagrangian is analytic at small momenta.

We now discuss the two remaining massless modes, these now being associated with density fluctuations. The first mode is one that would exist in a pure $^3$P$_2$ ( and also a $^1$S$_0$) neutron superfluid and it corresponds to the fluctuations of $\phi$ - the overall isotropic  phase of the condensate.  The other mode is related to density fluctuations of proton condensate + the electron gas and is denoted by the scalar field $\xi$. The general low energy effective field theory of these scalar modes is well studied \cite{Son:2005rv,Pethick:2010zf,Cirigliano:2011tj} and the low energy Largrangian density is given by 
\begin{eqnarray}
\mathcal{L} _{\rm phn} &=&  \frac{1}{2}(\partial_0\phi)^2 - \frac{v_n^2 }{2}(\partial_i\phi)^2+  \frac{1}{2}(\partial_0\xi)^2 - \frac{v_p^2 }{2}(\partial_i\xi)^2 \nn\\
&+&  v^2_{\rm np} ~\partial_0 \phi~\partial_0 \xi
+ \frac{1}{f_{\rm ep}} ~\partial_0 \xi ~\psi_{\rm e}^\dagger ~\psi_{\rm e}+ \cdots~
\label{eq:ep-Lag}\;, 
\end{eqnarray}
where we have also included the coupling to the electron field $\psi_e$. The coefficients of the leading order terms in the derivative expansion are related to simple thermodynamic derivates and can be obtained from the equation of state. They are given by 
\beq
v^2 = \frac{n_n}{m}~E_{nn}\,, \quad v_p^2 = \frac{n_p}{m}~E_{pp}\,, \quad v_{np}^2=  \frac{1}{2} ~\frac{k^2_{\rm Fp}}{\pi^2}\sqrt{\frac{k_{\rm Fn}}{k_{\rm Fp}}}~E_{\rm np}
\eeq
where $E_{\rm ij}=\partial^2 E(n_n,n_p)/(\partial n_i \partial n_j)$  and $E(n_n,n_p)$ is the energy density of the neutron-proton system.
The effective coupling between phonons in the $ep$ system and electron-hole states is calculated as in the jellium model and is given by $f_{\rm ep}= \sqrt{m_p~k_{\rm Fp}/\pi^2}$ \cite{FetterWalecka}. 
$E_{\rm np}$ arises solely due to nucleon-nucleon interactions and its value depends on the density, the equilibrium proton fraction and the equation of state model chosen. The low energy constants calculated using a representative microscopic equation of state from \cite{Gandolfi:2009nq} and the eigenmode velocities in units of the speed of light are shown in Table~\ref{table:vnp}.   

\begin{table}[t] 
\centering 
\begin{tabular}{c | c c c c c} 
\hline\hline 
$n_n$ (fm$^{-3}$) & 0.08  & 0.16 & 0.20 & 0.24 & 0.32 \\ 
\hline 
$x_p$ & 0.024  & 0.043  & 0.050  &  0.057 & 0.070\\  
\hline
$v_p^2$ & 0.029 & 0.049 & 0.060 & 0.072 &  0.104 \\    
\hline 
$v_n^2$ & 0.015  & 0.070 &   0.128 &  0.210 & 0.430 \\    
\hline
$v^2_{\rm np}$ & -0.034 & -0.016 & 0.024 & 0.086 & 0.268 \\
\hline
$v_1$ & 0.12 & 0.21 & 0.23 & 0.25  & 0.28  \\
\hline
$v_2$ & 0.17 & 0.26 & 0.36 & 0.46  & 0.71  \\
\hline
\end{tabular} 
\caption{Ambient conditions, low energy constants and eigenmode velocities $v_1$ and $v_2$ in units of the velocity of light for the equation of state from \cite{Gandolfi:2009nq} }
\label{table:vnp} 
\end{table}



The propagation of {\em angulons} and {\em superfluid phonons} can be damped by several processes. In the the following we estimate the mean free paths of phonons and angulons at low temperature $k_B T \ll \Delta$ to find that dominant decay mechanism is due to the excitation of electron-hole states.  First, we analyze the mean free paths of the two longitudinal {\em superfluid phonons}. In the absence of any mixing between these modes the $e-p$ mode couples strongly to the electron-hole excitations and its damping rate and the mean free paths are given by 
\bea
\Gamma_{\rm ep}(\omega =v_p~q)&=&\frac{3\pi}{2}~v_p^3~q, \\
 \lambda_{\rm ep}(\omega =v_p~q)&=&\frac{c}{\Gamma_{\rm ep}(\omega)} 
\simeq 1.4\times10^{-9}~\left[\frac{10~{\rm keV}}{\omega}\right] \left[\frac{0.3}{v_p}\right]~{\rm cm}  \,, \nn
\label{eq:ep}
\eea 
respectively. The thermal average mean free path is well defined and is given by $ \langle \lambda_{\rm ep}(T) \rangle= \pi/(18\zeta [3]~v_p~T) \approx 10^{-9}~(0.3/v_p)~T_8^{-1}~{\rm cm}$, where $T_8$ is the temperature measured in units of $10^8$ K.  

$v_{\rm np}$ mixes the the proton-electron mode with the neutron {\em superfluid phonon} mode. We find that both eigenmodes decay predominantly by coupling to electron-hole excitations (Landau damping). This mixing is similar to the mixing between the longitudinal phonons of the nuclear lattice and the neutron {\em superfluid phonons} in the inner crust of the neutron star \cite{Cirigliano:2011tj}.  The velocity and damping rates of the two longitudinal eigenmodes can be obtained as solutions to the equation
\beq
(\omega^2 - v_n^2 q^2)(\omega^2 - v_p^2 q^2 -2i~\omega~\Gamma_{\rm e-p}(\omega))-2 v^4_{\rm np} \omega^4=0
\eeq
In the limit of weak mixing the scattering rate of the predominantly $ep$-mode is $\approx \Gamma_{\rm ep}(\omega =v_p~q)$ given in \eq{eq:ep}, and the scattering rate of the predominantly neutron superfluid mode is 
\beq 
\Gamma_{\rm \phi}(\omega =v_n~q)\approx \frac{v^4_{\rm np}~\Gamma_{\rm ep}(\omega)}{(1-(v_p/v_n)^2)^2+9\pi^2 ~v_p^4}\,,
\eeq 
and when $v_n \gg v_p$ and $v_p \ll 1$, $\Gamma_{\rm \phi}(\omega =v~q) \approx v^4_{\rm np} \Gamma_{\rm ep}(\omega)$. Since typical values of $v^4_{\rm np}$ are in the range  $10^{-4}-10^{-1}$, we can conclude that the mean free path of the predominantly neutron superfluid mode will be in the range $\lambda_{\phi}\approx 10^{-5}(0.3/v)~T_8^{-1}$ cm to $\lambda_{\phi}\approx 10^{-8}(0.3/v)~T_8^{-1}$ cm. Although they are typically much larger than those corresponding to $ep$ mode,  as we shall see later, these are still too small to contribute significantly to any transport phenomena. 

\begin{figure}[t]
\centerline{\includegraphics[width=8cm]{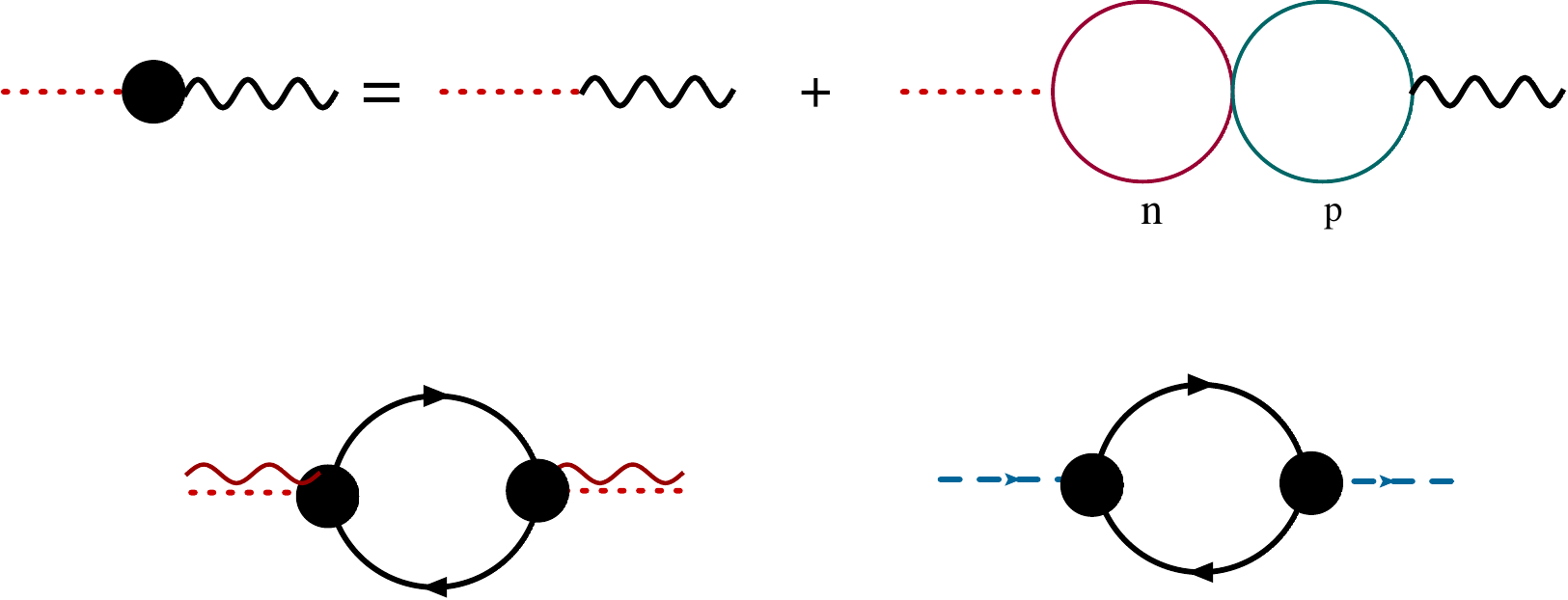}}
\caption{The top line shows the  magnetic moment and proton mediated mixing processes, respectively. Angulon, photon, electron, neutron and proton propagators are shown as a dotted, wavy and solid black, solid red and solid blue lines lines, respectively and nucleon loops include both normal and anomalous diagrams. The lower graphs contribute to the imaginary part of the self-energy of the angulon-magnetic photon mixed mode (left) and the electron-proton-neutron phonon mode (right).  }
\label{fig:diagrams}
\end{figure}  
We now turn to the calculation of the {\em angulon} mean free path.The angulon-angulon scattering amplitude is $\propto p^2/f_\beta^2$ since the powers of $p$ are fixed by dimensional analysis. Its contribution to the mean free path can then be easily estimated and we find $\lambda_{\rm ang-ang} \approx v_\beta^3 f_\beta^4/T^5$. For $T\lesssim10^9 $ K, $\lambda_{\rm ang-ang}\gg R$ where $R \simeq 10$ km is the radius of the neutron star, and implies that angulon-angulon processes are irrelevant. 
%
%

Angulons mix with the magnetic photons due to two processes. One mixing mechanism is due to the magnetic moment of the neutron and is described by the lagrangian in \eq{eq:L_ang}  the other is mediated by protons which, as charged particles, couple to photons. These two processes are depicted in Fig.~\ref{fig:diagrams}. The latter indirect coupling necessarily involves a spin flip of both neutrons (on account of form  the angulon-neutron coupling) and protons. Thus, only the {\em magnetic} photon mixes with the angulon and this mixing is suppressed by a power of the proton velocity change $\sim p/M$, the same suppression appearing in the magnetic moment process. 
We find that the proton mediated mixing is smaller than the mixing generated by the neutron magnetic moment. For the estimates we present here, we will neglect the proton mediated mixing. 

Since magnetic photons are damped by electron-hole excitations, mixing ensures that angulons are also damped. 
The angulon scattering rate off electrons is given by 
\bea
\Gamma_{\rm ang}(\omega,q) \simeq  \frac{1}{\omega}\left[ \frac{e f_\beta~g_n }{2M}~q~ \omega\right]^2~{\rm Im} ~D(\omega,q) \,,
\eea
where $D(\omega,q) = (\omega^2-q^2- \omega^2_P- i q^2_{\rm TFe}~ (\omega/q))^{-1}$ is the dressed  photon propagator, and  $q_{\rm TFe} = \sqrt{4 \pi e^2 ~ \partial n_e/\partial \mu_e} $ is the electron Debye screening momentum, and $\omega_P=\sqrt{4 \pi e^2 ~ n_p/m} $ is the proton plasma frequency.  
Since $ q^2_{\rm TFe} \gg \omega^2_P$ we can write
\bea
\Gamma_{\rm ang} (\omega,q) \simeq   e^2 f^2_\beta ~ \frac{g^2_n }{4M^2~q^2_{\rm TFe}}~q^3 = \frac{\pi f^2_\beta ~ g^2_n }{16M^2~k^2_{\rm Fe}}~q^3  \,,
\label{eq:Gamma_ang}
\eea
From the angulon width estimated above we can determine the angulon mean free path 
\bea
\lambda_{\rm ang}(\omega) &=& \frac{v_\beta}{\Gamma(\omega = v_\beta q)} =\frac{16M^2~k^2_{\rm Fe}v^4_\beta}{\pi f^2_\beta ~ g^2_n }~\frac{1}{\omega^3}\,, \\
\label{eq:lambda_ang}
&\approx & 1.7  \gamma^4 ~ v^3_F \left[\frac{k_{\rm Fe}}{100 ~{\rm MeV}} \right]^2  \left[ \frac{10~{\rm keV}}{\omega} \right]^3 {\rm cm} \,, 
\eea 
where $\gamma= v_\beta / v_F$ and $v_\beta$ is the mean velocity of the angulon. 

Transport properties like the heat conductivity $\kappa$ and the shear viscosity $\eta$ can be computed  by solving the Boltzmann equation for phonons and angulons now that we have identified relevant scattering processes. To obtain simple estimates we use results from kinetic theory. In kinetic theory, the thermal conductivity and shear viscosity are given by  $\kappa \sim(1/3) C_V v  \langle \lambda \rangle $, $\eta \sim(1/3) n  \langle p\rangle  \langle \lambda \rangle $, respectively, 
where $C_V \simeq 2 \pi^2~T^3/(15 v^3)$ is the specific heat, $n \simeq \zeta[3]T^3/(\pi^2 v^3)$ is the number density,  $v$ is the velocity, $\langle p\rangle \simeq T/v$ is the average thermal momentum, and $\langle \lambda \rangle $ a thermally averaged mean free path of the phonon/angulon. 

In cgs units the phonon/angulon contribution can be written as 
\beq
\kappa_{\rm phn/ang} = 1.7\times 10^{21} ~T_8^3~\left(\frac{0.3}{v}\right)^3~ \left(\frac{\langle \lambda \rangle}{\rm cm}\right) \frac{\rm erg}{\rm cm~s~K}
\eeq
The electron contribution to thermal conductivity has been calculated in earlier work and was found to be in the range $10^{22}-10^{24}$ erg/cm/s/K for typical ambient conditions in the neutron star core  \cite{Shternin:2007ee}. Since  $  \langle \lambda_{\rm ep} \rangle \ll \langle \lambda_\phi \rangle \lesssim 10^{-5} $ cm, we can safely neglect the contribution from both longitudinal phonons to thermal conductivity. 

Estimating the angulon contribtuion is bit trickier because $\lambda_{\rm ang}(\omega) \propto 1/\omega^3$ and the naive thermal average  mean free path diverges. Here it is appropriate to write the thermal conductivity as     
\bea
\kappa_{\rm ang} &=& \frac{1}{3}\int \frac{d^3 q}{(2\pi)^3} \frac{d}{dT}\left(  \frac{\bar{v}q}{e^{\beta \bar{v} q}-1}\right)  \bar{v}~ \lambda_{\rm ang}(q)\nn\\
&\simeq&
\frac{8 ~\gamma^2~v_F~k_{\rm Fe}^2}{3\pi~g_n^2} \int_{\bar x}^\infty \frac{x~e^x}{(e^{x}-1)^2}  \\
&\simeq & 2.5 \times 10^{19}~\left(\frac{k_{\rm Fe}}{100~{\rm MeV}}\right)^2 \gamma^2~v_F~(1-\ln{(\bar{x})}) \frac{\rm ergs}{\rm cm~s~K}
\nonumber 
\label{eq:heat}
\eea 
where $\beta =1/k_BT$ and $\bar{v}$ is the angle average velocity. The lower limit  $\bar{x}=  \beta \bar{v} q_c \ll 1 $ is introduced because the Bose distribution function is meaningful for low energy angulons only when the mean free path $\lambda_{\rm ang}(\omega =  \bar{v} q_c) \ll  R_c$ where $R_c \simeq 5-10$ kms is radius of the core. For relevant neutron star temperatures we find that $\bar{x}$ is in the range $10^{-4}-10^{-3}$. $\kappa_{\rm ang}$ is nearly independent of temperature and
for typical values $\gamma \simeq 0.5-0.7$, $v_F\simeq 1/3$ and $ \bar{x}\simeq 10^{-3}$, 
$\kappa_{\rm ang} \approx 1.6 \times 10^{19}$ ergs/(cm\ s\ K). This is a few orders of magnitude smaller than earlier estimates of the electron thermal conductivity. Similarly, an estimate of the shear viscosity shows that it too is small compared to the electronic contribution. The elementary process between electrons and GBs could nevertheless play role in coupling the dynamics of the multicomponent core of the neutron star. In addition, long wavelength modes with $q < q_c$ are ballistic, and despite their low production rates, could transport energy and momentum and warrants further study.

Perhaps the most consequential finding is the long lifetime of the long wavelength angulon and phonon modes.  This implies that they would play a key role in neutron star seismology. If excited these modes will have different characteristic frequencies given by $\Omega = 2\pi v/R$ where $v$ is the phonon/angulon velocity, and different damping timescale $ \tau = \lambda(q\simeq 2 \pi/R)/v$ where $\lambda$ is the corresponding mean free path. This is likely to produce a unique and discernible  spectrum and time evolution. The low frequency longitudinal mode will damp quickly in the bulk, the superfluid phonon modes will damp on a longer time scale and the angulon modes will likely be damped only at the crust-core interface as they are essentially undamped in bulk. The long decay constant of angulons is a unique property among oscillations modes of neutron stars and it can be traced back to the fact that angulon interactions with electrons is mediated by magnetic forces suppressed by powers of $q/M$. Further, the anisotropy of the angulon modes, and that they couple to the magnetic field suggests unique observable consequences that need to be explored. For example, catastrophic processes such as giant flares or tidal perturbations that occur prior to binary neutron star mergers  could trigger seismic activity in the core leading to coupled dynamics of the angulon fields and the large scale magnetic field anchored in the star.     

We hope that our calculation of the mode velocities and damping rates for the elementary excitations with macroscopic and microscopic wavelengths in the neutron star core will motivate further study. There are several issues that warrant a more detailed analysis. These include the role of ballistic angulon modes, dissipation of long wavelength angulons at the crust-core boundary, and the 
role of phonon-emission and absorption reactions on the electronic transport properties.               
\begin{acknowledgments}
The work of S.R. was supported by the DOE Grant No. DE-FG02-00ER41132 and by the 
Topical Collaboration to study {\it Neutrinos and nucleosynthesis in hot and dense matter}.    
The work of P.B. was supported by the DOE Grant No. DEFG02-93ER-40762.   
\end{acknowledgments}

\bibliographystyle{apsrev4-1}
\bibliography{coremodes}

\end{document}